\documentclass{amsart}

\usepackage{amsmath,amscd,amssymb,amsthm}
\usepackage{latexsym,enumerate,graphicx,rotating,braket,mathrsfs,url}
\usepackage{hyperref}
\usepackage{geometry}
\newcommand{\bibfilename}{unibib}
\numberwithin{equation}{section}

\theoremstyle{plain}
\newtheorem{thm_}[equation]{Theorem}
\newtheorem{lemma_}[equation]{Lemma}
\newtheorem{prop_}[equation]{Proposition}
\newtheorem{cor_}[equation]{Corollary}
\newtheorem{eg_}[equation]{Example}
\newtheorem{con_}[equation]{Conjecture}
\newtheorem*{cons_}{Conjecture}

\theoremstyle{definition}
\newtheorem{thmu_}[equation]{Theorem}
\newtheorem*{thmus_}{Theorem}
\newtheorem{propu_}[equation]{Proposition}
\newtheorem*{propus_}{Proposition}
\newtheorem{coru_}[equation]{Corollary}
\newtheorem{lemu_}[equation]{Lemma}
\newtheorem{egu_}[equation]{Example}
\newtheorem*{egus_}{Example}
\newtheorem{def_}[equation]{Definition}
\newtheorem*{defs_}{Definition}

\theoremstyle{remark}
\newtheorem{rk_}[equation]{Remark}

\newcommand{\thm}[1]{\begin{thm_}#1\end{thm_}}
\newcommand{\thmu}[1]{\begin{thmu_}#1\end{thmu_}}

\newcommand{\lemm}[1]{\begin{lemma_}#1\end{lemma_}}
\newcommand{\lemu}[1]{\begin{lemu_}#1\end{lemu_}}
\newcommand{\eg}[1]{\begin{eg_}#1\end{eg_}}

\newcommand{\prop}[1]{\begin{prop_}#1\end{prop_}}
\newcommand{\propu}[1]{\begin{propu_}#1\end{propu_}}

\newcommand{\defi}[1]{\begin{def_}#1\end{def_}}

\newcommand{\rk}[1]{\begin{rk_}#1\end{rk_}}
\newcommand{\cor}[1]{\begin{cor_}#1\end{cor_}}

\newcommand{\pf}[1]{\begin{proof}#1\end{proof}}
\newcommand{\pfs}[1]{\begin{proof}[Proof~$($sketch$)$]#1\end{proof}}
\DeclareMathOperator{\ord}{ord}
\DeclareMathOperator{\GL}{GL}
\DeclareMathOperator{\Gal}{Gal}

\newcommand{\fracn}[2]{\genfrac{(}{)}{}{}{#1}{#2}}

\newcommand{\ZZ}{\mathbb Z}
\newcommand{\QQ}{\mathbb Q}
\newcommand{\RR}{\mathbb R}
\newcommand{\CC}{\mathbb C}

\renewcommand{\a}{\mathfrak a}

\renewcommand{\o}{\mathfrak o}
\newcommand{\p}{\mathfrak p}
\renewcommand{\P}{\mathfrak P}

\newcommand{\Q}{\mathfrak Q}
\newcommand{\q}{\mathfrak q}

\renewcommand{\t}{\times}

\newcommand{\lr}{\longrightarrow}

\newcommand{\s}{\sigma}

\newcommand{\eq}[1]{\begin{equation}#1\end{equation}}
\newcommand{\eqn}[1]{\begin{equation*}#1\end{equation*}}
\newcommand{\ga}[1]{\begin{gather}#1\end{gather}}
\newcommand{\gan}[1]{\begin{gather*}#1\end{gather*}}

\newcommand{\aln}[1]{\begin{align*}#1\end{align*}}
\renewcommand{\sp}[1]{\begin{split}#1\end{split}}

\newcommand{\itmz}[1]{\begin{itemize}#1\end{itemize}}
\newcommand{\enmt}[1]{\begin{enumerate}#1\end{enumerate}}
\renewcommand{\it}{\item}

\newcommand{\itm}[1]{\it[\upshape{(#1)}]}
\newcommand{\newnoindbf}[1]{\vspace{2mm}\noindent\textbf{#1}}

\begin{document}
%title & author
\title[Non-existence of PPSs and PAPSs]{On the Non-existence of certain classes of perfect $p$-ary sequences and
perfect almost $p$-ary sequences}
\author[C. Lv]{Chang Lv}
\address{
State Key Laboratory Of Information Security, Institute of Information Engineering, Chinese Academy of Sciences, Beijing 100093, P.R. China
}
\email{lvchang@iie.ac.cn}


\date{\today}
%\thanks{This work was supported by xxxxxxxxxxxxxxx.
%}
\begin{abstract}
We obtain  new non-existence results of perfect $p$-ary sequences with period $n$ (called type $[p, n]$).
The first case is a class with type  $[p\equiv5\pmod 8,p^aqn']$.
The second case contains five types  $[p\equiv3\pmod 4,p^aq^ln']$ for certain $p, q$ and $l$.
Moreover, we also have similar non-existence results for PAPSs.
\end{abstract}
\maketitle

\section{Introduction}\label{sec_intro}
Let $n$ be a positive integer, $p$ a rational prime   and $\zeta_p$
a   primitive $p$-th root of unity (we can take $\zeta_p$ to be $\exp(\frac{2\pi i}{p})$).
\newcommand{\bfa}{\mathbf a}
\defi{\label{def_pps}
 A  complex sequence $\bfa=(a_0, a_1,\dots,a_{n-1},\dots)$ 
 with period $n$
is called a \emph{$p$-ary sequence} (resp. an \emph{almost $p$-ary sequence}) if
$a_j=\zeta_p^{b_j}$ where $b_j\in\ZZ$ for all $i\ge0$
(resp. $a_0=0$  and
$a_j=\zeta_p^{b_j}$ where $b_j\in\ZZ$ for all $1\le  i \le n-1$).

 A  complex sequence $\bfa=(a_0, a_1,\dots,a_{n-1},\dots)$ 
 with period $n$
is called \emph{perfect} if 
$C_\bfa(t)=0$ for all $1\le t\le n-1$,
where \eqn{
C_\bfa(t)=\sum_{k=0}^{n-1}a_k\bar a_{k+t}
} is the \emph{autocorrelation} with a bar meaning the complex conjugation.

For simplicity, we  denote a perfect $p$-ary 
(resp. an perfect almost $p$-ary) sequence with period $n$ 
as a \emph{PPS} (resp. \emph{PAPS})  with  \emph{type} $[p, n]$. 
}
%PPSs have been used in many fields such as  difference sets, coding theory, cryptography and sequence designs.
%For more background information and its applications we refer the reader to \cite{Dillon,OSW,Kumar}.

A natural question is when PPSs (PAPSs) do exist. 
This is equivalent to the existence of certain kinds of relative difference sets.
See \cite{chee2010almost,jungnickel1999perfect,liu2016new} for details.
Their results imply that PPSs (PAPSs) can be constructed if the
 corresponding relative difference sets exist. 
By using various techniques in combinatorial design
theory, several classes of such sequences have been constructed 
(see \cite{chee2010almost,jungnickel1999perfect,liu2016new,krengel2010some}).
On the other hand, there are some nonexistence results on such sequences 
(and related difference sets), see
 \cite{chee2010almost,jungnickel1999perfect,ma2009non,ozbudak2012nonexistence}.
Here we need the concept of ``self-conjugate". See \cite{Washington,liu2016new}.
\defi{
Let $p$ be a prime integer, $m = p^a m'$ where $a\ge 0$ and $(p, m') = 1$. We
call $p$ to be \emph{self-conjugated} with respect to $m$ if 
there exists $s \in\ZZ$ such that $p^s \equiv −1\pmod {m'}$.
Namely, if $-1\in\left<p\right>\subseteq(\ZZ/m'\ZZ)^\t$.
}
Now we give a list of typical non-existence results of PPSs (PAPSs) with reference at the beginning of each item:
\enmt{
\it (Ma and Ng \cite{ma2009non})\label{it_ma} PPSs with type $[p, q^ln']$ where $p\neq q$ are two primes,
 $p\ge3,\ (q,n')=1,\ q$ is self-conjugate w.r.t. $p$ and  $l\ge1$ is odd.
\it (Liu and Feng \cite{liu2016new})\label{it_liu} PPSs with type $[p, p^aq^ln']$ where $p\equiv3\pmod4$ is a prime,
$q$ is another prime with $(q-1, p) = 1,\ \fracn{q}{p}=1$ and  $\ord_p(q)$ being odd,
 $n'$ satisfies that $n'=1$ or $\fracn{p'}{p}=-1$ for all prime divisor $p'$ of $n'$,
$a\ge1$ and $l$ is odd such that $ l < \lambda/s$ where $s = (p-1)/\ord_p(q)$ and $\lambda$
 is the smallest odd integer such that $x^2 + py^2 = 4q^\lambda$ has
solution $(x, y),\ x, y \in\ZZ$.
\it (Liu and Feng \cite{liu2016new}) PAPSs with type $[p, q^ln'+1]$
 where $p\equiv3\pmod4,\ p\mid q^ln'-1$ is a prime,
$q,n',a$ and $l$ are the same as the above \eqref{it_liu}.
}

In this article we have two main results.
The first one shows the non-existence of PPSs with type $[p,p^aqn']$,
 where $p\equiv5\pmod8$ is a prime, $q$ runs through a infinite set of primes and $n'$ is the same
as  \eqref{it_liu} in the  above list.
\thm{ \label{thm_p5mod8}
Let $p\equiv5\pmod 8$ be a prime and $\tilde Q_p=\set{q\text{ is a prime } | \ord_p(q)=(p-1)/4}$.
Then there exists a lower bound $p_0$, and an infinite set $Q_p\subseteq \tilde Q_p$
for each $p$, such that
if $p>p_0$, there is no PPSs with type $[p, n=p^aqn']$ for all integers $a\ge1$, $q\in Q_p$ and $n'$ 
such that $n'=1$ or $\fracn{p'}{p}=-1$ for all prime divisor $p'$ of $n'$.
}
\rk{\label{rk_p5mod8}
Since  $p\equiv5\pmod 8$,  we have   $\ord_p(q)=(p-1)/4$  is odd for all $q\in Q_p$.
It follows that $q$ is not self-conjugate w.r.t $p$,
which says that our case is not contained in 
\cite{ma2009non} (\eqref{it_ma} in the above list).
Moreover, our case is also different from
\cite{liu2016new} (\eqref{it_liu} in the above list)
since $p\not\equiv3\pmod4$.
Thus  our result is new.
}

In the second main result  we   obtain  the non-existence  of PPSs with five types:
\thm{ \label{thm_sp}
Let $p\equiv3$ be a prime, $q\neq p$ another prime  and $f=\ord_p(q)$.
Suppose that the triple $(p, f, l_0)$ equals to one of the following value:
\eqn{
(31 , 5 , 1),\ (127, 9 , 1),\ (127, 21, 3),\ (139, 23, 1),\ (151, 15, 3).
}
Define \aln{
\Xi_{31}(x)					&=  x^3 + x - 1,\\
\Xi_{127}(x)					&=  x^5 - x^4 - 2x^3 + x^2 + 3x - 1,\\
\Xi_{139}(x)					&=  x^3 - x^2 + x + 2,\\
\text{and}\quad \Xi_{151}(x)	&=  x^7 - x^6 + x^5 + 3x^3 - x^2 + 3x + 1.
}
Suppose further that for each $p\in\set{31,127,139,151}$, the corresponding $q$ satisfies that
$\Xi_p(x)\equiv0\pmod q$ is not  solvable.
Then there is no PPPs with type $[p, n=p^aq^ln']$ for all integers $a\ge1$, $l$ odd, $1\le l\le l_0$ 
 and $n'$ such that $n'=1$ or $\fracn{p'}{p}=-1$ for all prime divisor $p'$ of $n'$.
}
\rk{
For the same reason, this case is also not contained in 
\cite{ma2009non} (\eqref{it_ma} in the above list),
and  the result \cite{liu2016new} (\eqref{it_liu} in the above list)
 can only deal with  type  $[p, p^aq^ln']$ 
where $ l < \lambda/s$.
By direct  calculation for $(p,f)$ in the cases listed in Theorem \ref{thm_sp},
the corresponding $\lambda/s\le l_0$.
 Thus the results in Theorem \ref{thm_sp} are also new.
}

For the proofs of the two theorems, we need some facts in algebraic number theory which are contained 
in Section \ref{sec_facts}. 
With these preparations, we can prove Theorem \ref{thm_p5mod8} and \ref{thm_sp}
in Section \ref{sec_p5mod8} and \ref{sec_sp}, respectively.

We also have corresponding non-existence results for PAPSs, which are similar to 
Theorem \ref{thm_p5mod8} and \ref{thm_sp}. See the last section.

\section{Basic Facts in Algebraic Number Theory}\label{sec_facts}
 The methods for proving non-existence results of PPSs  often involve algebraic number theory,
  mainly the basic arithmetic (ideals, units, class groups etc.) of cyclotomic fields and their subfields.
  The standard reference are \cite{janusz}  and \cite{Washington}. In this section, we list some facts needed later, 
with proofs or references. 
The reader who does not care the proofs may skip to the next section.

For any number field  $F$, denote by $\o_F$ the ring of integers of $F$.
The latter ring is a Dedekind domain and we often consider the \emph{fractional ideals} in it,
which are $\o_F$ modules of the form $\a/\alpha$, where $\a\subseteq \o_F$ is an integral ideal and 
$\alpha\in\o_F$ is a nonzero element.
Denote by $I_F$  the set of nonzero fractional ideals of $F$,
which, one can show, under multiplication, is a free abelian group
generated by all prime ideals.
By a principal fractional ideal we mean a fractional ideal of the form $\alpha\o_F$ where $\alpha\in F^\t$.
Clearly,  $P_F\subseteq I_F$ as a subgroup,
and the quotient $I_F/P_F$, denoted by $Cl(F)$,
is called the  \emph{class group} of $F$.
Class groups play an important role in classical algebraic number theory.
One of the nontrivial facts is that $Cl(F)$ is a finite abelian group for all $F$, 
and  by $h(F)$ we denote the cardinality of $Cl(F)$, called the \emph{class number} of $F$.

We need the basic knowledge of the decompositions of prime ideals in  extension fields,
the  decomposition  groups and  the decomposition fields. We refer the reader to 
\cite[Section I.6, Section III.7]{janusz}. 
We also use properties of Artin maps and we need a corollary of class field theory,
that is, there exists a finite unramified abelian extension $H_F/F$ 
(called the \emph{Hilbert class field})
for every $F$, such that  the map
$Cl(F)\lr\Gal(H_F/F)$ induced by Artin map 
is an isomorphism (see \cite[Section V.13]{janusz}).
In particular, a prime ideal $\p$ of $F$ is principal if and only if  $\p$ splits completely
in $H_F$, and we have $h(F)=[H_F:F]$.

For two subfields of the cyclotomic field $\QQ(\zeta_{p^e})$ where $p^e$ is a prime power,
we have the divisibility of class numbers.
\lemm{\label{lem_h_div}
Let  $L=\QQ(\zeta_{p^e})$ and $F\subseteq E\subseteq L$ be two subfields of $L$. 
Then we have $h(F)\mid h(E)$.
}
\pf{
Since $E/F$ is abelian and $p$ is totally ramified in $L/\QQ$, the result follows from 
 \cite[Proposition~4.11]{Washington}.
}
\lemm{\label{lem_cl_inj}
Let $E/F$ be two  number fields.
Then  the canonical morphism $j_{E/F}: Cl(F)\lr Cl(E)$ sending 
$\a$ to $\a\o_E$ is injective, provided that $\gcd(h(F),[E:F])=1$.
}
\pf{
The argument is quite  simple.
Let $\a$ be a fractional ideal of $F$ such that $\a\o_E$ is trivial in $Cl(E)$.
Then $\a\o_E=\alpha\o_E$ for some $\alpha\in E$. Taking norm
 to $F$ gives $\a^{[E:F]}=N_{E/F}(\alpha)\o_F$.
But $\gcd(h(F),[E:F])=1$, therefore  raising to the power to $[E:F]$ 
 is an automorphism on $Cl(F)$. 
Hence $\a$ is also trivial in $Cl(F)$. This prove the injectivity.
}
For some cases, we have the following   more strong  statements.
\prop{\label{prop_cl_inj}
Let $p\equiv3\pmod4,\ p>3$ be a prime and $L=\QQ(\zeta_{p^e}),\ F:=\QQ(\sqrt{-p})$.
It is well-known that  $F$ is a subfield of $L$.
Let $E$ be any number field such that $F\subseteq E\subseteq L$.
Then $j_{E/F}: Cl(F)\lr Cl(E)$ is injective.
}
\pf{
The statement of \cite[Corollary to Proposition 4, pp. 2723]{schoof2010visibility} says that
if $M$ is any subfield of $\QQ(\zeta_n)$ with the only roots of unity $\pm1$, and 
$\a$ is an ideal of $M$ such that $\a\bar\a$ is principal in $M$ and
 $\a$ is principal in  $\QQ(\zeta_n)$, 
then $\a^4$ is principal in $M$.
Now we apply this result with $M=F$ and  $n=p$. 
Let $\a$ be any ideal of $F$ that is principal in $E$. 
Then  $\a$ become principal in $L$.
Also $\a\bar\a$ is clearly principal in $F$  since $F$ is imaginary quadratic.
It follows that $\a^4$ is principal in $F$.
On the other hand, by  Gauss' genus theory (c.f. \cite[Theorem 10.4 (b)]{Washington}) 
 or Lemma \ref{lem_cm_parity} below,
we know that $h(F)$ is odd.
Thus $\a$ is principal in $F$.
The injectivity follows.
}

To show that the set $Q_p$ 
in Theorem \ref{thm_p5mod8} 
 is infinite in the subsequent section,
 we  need a special case of Chebotarev's density theorem and compare class numbers.
We first introduce
\defi{
Let $K$ be any number field and  $S$ be a set of prime ideals of $\o_K$.
Denote all prime ideals of $\o_K$ by $\mathcal P_K$.
The \emph{Dirichlet density} of $S$ is the limit (if exists)
\eqn{
\delta=\lim_{s\rightarrow1^+} 
	\frac{\sum_{\p\in S} \frac{1}{N_{K/\QQ}(\p)^s}}{\sum_{\p\in \mathcal P_K} \frac{1}{N_{K/\QQ}(\p)^s}},
}
denoted as  $\delta(S)=\delta$.
}
There may exists some other definitions but they are equivalent.
Now we have  the statement:
\prop{\label{prop_density}
Let $L/K$ be abelian extension of two number fields with Galois group $G$ and fix an
element $\s\in G$.
Let $S$ be the set of prime  ideal $\p$ of $K$ whose Artin map $(\p,L/K)$ is   $\s$.
 Then $S$ has Dirichlet density $\delta(S)=1/\#G$.
}
\pf{See, for example \cite[Theorem $13.4$]{neukirch_alnt}.
}

Next we  consider a wider class of number fields containing cyclotomic fields, namely:
\defi{
A \emph{CM-field} $E$ is a totally imaginary quadratic extension of a totally real number field $E^+$.
The field $E^+$ is  the \emph{maximal real subfield} of $E$.
That a field is totally real (resp. imaginary) means that all embeddings of the field into $\CC$ is real (resp. imaginary).
}
As mentioned above, we want to compare certain class numbers. 
For this purpose, we mainly use the following facts about CM-fields:
\propu{[c.f. \cite{Washington}, Section $4$, pp. $38$-$43$] \label{prop_cm}
Let $E$ be CM and $E^+$ its maximal real subfield.
For convenience, let $h, U, W, R$ and $d$ be the  class number, 
unit group, group of roots of unity, regulator and discriminant of $E$ respectively,
and let  $h^+, U^+, R^+$ and $d^+$ denote the corresponding objects for $E^+$. Then
we have:
\itmz{
\itm{a} The class number $h^+$ divides $h$, and the quotient $h^-$ is called the \emph{relative class number}.
\itm{b} The index $Q:=[U:WE^+]=1\text{ or }2$.
\itm{c} The quotient $R/R^+=\frac{1}{Q}2^r$, where $r:=\frac{1}{2}[E:\QQ]-1$.
\itm{d} (Brauer-Siegel theorem) Suppose $E$ runs through a sequence of number fields normal over $\QQ$ 
(not necessary CM) such that \eqn{
\frac{[E:\QQ]}{\log|d_E|}\rightarrow 0.}
Then \eqn{
\frac{\log(h(E) R_E)}{\log\sqrt{|d_E|}}\rightarrow 1.}
}}
We also need a result for the parity of the class numbers of a special class of  CM-fields.
\lemu{[See \cite{conner1988class}, Corollary 13.13]\label{lem_cm_parity}
Let $E$ be CM which is Galois over $\QQ$ with  $\Gal(E/\QQ)$ a cyclic group of order $2^k,\ k\ge1$.
Then $h(E)$ is odd if  and only if  exactly one finite rational prime ramifies in $E/\QQ$.
}

\bigskip
Next we introduce  Stickelberger ideals.
Suppose $p$ is  a prime, $K=\QQ(\zeta_p)$ 
and $G=\Gal(K/\QQ)\cong(\ZZ/p\ZZ)^\t$.
\defi{
The \emph{Stickelberger element} $\theta=\theta_p\in\QQ[G]$ 
is defined by \eqn{
\theta=\sum_{a\in(\ZZ/p\ZZ)^\t}\left\{\frac{a}{p}\right\}\s_a^{-1}
} where $\{\frac{a}{p}\} = \frac{a}{p}- [\frac{a}{p}]$,
and the \emph{Stickelberger ideal} $S_p$ of $\ZZ[G]$ is defined by \eqn{
S_p=\ZZ[G]\theta\cap\ZZ[G].
}
}
We mainly use these following properties of the Stickelberger ideal:
\prop{\label{prop_stick}
We have: 
\itmz{
\itm{a} For $(c, p) = 1$, the element $(c-\s_c)\theta$ are in $S_p$.
\itm{b} The Stickelberger ideal $S_p$ annihilates the ideal class group $Cl(M)$,
where $M$ is a subfield of $K$ such that  $p$ is the minimal integer with the property that
$M\subseteq \QQ(\zeta_p)$.
}}
\pf{ See \cite[Lemma 6.9 and Theorem 6.10]{Washington}.}

\newnoindbf{Notation.}  Through this  paper, we fix the following notation.
Let $p$ be an odd prime and denote   $\zeta_k$ a  primitive $k$-th root of unity.
Let $K=\QQ(\zeta_p)$.
In the remaining of this paper we mainly deal with $\QQ(\zeta_p)$ and write $\zeta = \zeta_p$ for
simplicity. 
Let $G=\Gal(K/\QQ)$. It's well-known that $G\cong(\ZZ/N\ZZ)^\t$, the isomorphism being 
$c\mapsto(\s_c: \zeta\mapsto \zeta^c)$ for $c\in(\ZZ/N\ZZ)^\t$.

The starting point of our method is the following 
\prop{\label{prop_eq_pps}
If there exist PPS  with type $[p, n]$, then $p\mid n$ and 
$\alpha\bar\alpha=n$ for some $\alpha\in\ZZ[\zeta_p]$.
}
\pf{
The result is obtained by applying different sets. 
See, for example, \cite[Theorem 1.4 (1)]{liu2016new} and the remarks after it.
}
Thus for our purpose we need to investigate the equation $\alpha\bar{\alpha}=n$ where $\alpha \in \ZZ[\zeta_p]$. 
So we mainly study the idealic behaviour of each $p$ dividing $n$, in the cyclotomic field $K$.

\section{Non-existence result for PPSs with type $[p\equiv5\pmod 8,p^aqn']$}\label{sec_p5mod8}
In this section, we will prove  Theorem \ref{thm_p5mod8}.
We start with the definition of $Q_p$.
As  the assumptions  in the theorem, let $p\equiv5\pmod 8$ be a prime and 
\eqn{
\tilde Q_p=\set{q\text{ is a prime } | \ord_p(q)=(p-1)/4}.
}
Let $q\in \tilde Q_p$ so  $\ord_p(q)=(p-1)/4$.
Let $K = \QQ(\zeta_p)$ and $E$ be the unique subfield of $K$ having degree $4$ over $\QQ$.
Then the order of $q$ modulo $p$ tells us that $E$ is the 
decomposition group of $q$ in $K$ and depends only in $p$.
Thus we write $E_p=E$ and it is well known that  $K$ contains the unique 
real quadratic subfield $F_p=\QQ(\sqrt{p})\subset E_p$

Actually one can define
\eq{\label{eq_qp}
Q_p = \left\{q \in\tilde Q_p  \Biggm| \sp{
	 &\text{ there is a prime ideal $\Q$ in $E_p$  lying over $q$ such that }\\
     &\text{ $\Q$ is not principal while  $\q=\P\cap \o_{F_p}$  is principal}
	} \right \}.
}
To  show that  $Q_p$  is infinite, we only need to show that the Dirichlet density
$\delta(Q_p)>0$, since any finite set has zero density by the definition. 
\lemm{\label{lem_density_prin}
Let $L/M$ be cyclic extension of  two number fields such that
they are both Galois over $\QQ$ and 
there is some finite prime in $M$ totally ramified in $L$. 
Define \eqn{
S(f, L/M)= \left\{p \text{ is a prime number } \Biggm| \sp{
	 &\text{ $p$ split completely in $M$ and there is a principal }\\ 
	 &\text{ prime ideal $\p$ in $M$ lying over $p$ and the order }\\
	 &\text{ of the Artin map $(\p, L/M)$ is $f$ }
	} \right \}.
}
Then we have \eqn{
\delta(S(f, L/M)) = \frac{\varphi(f)}{[L:\QQ]\ h(M)},
}
with $\varphi$ being the Euler's totient function.
}
\pf{
Let $H_M$ be the Hilbert class field of $M$. Since there is a finite prime 
totally ramified in $L/M$ and $H_M/M$ is unramified, we have $H_M\cap L=M$.
Hence we have a natural isomorphism 
\eq{\label{eq_gal_split}
\Gal(LH_M/M)\cong\Gal(L/M)\t\Gal(H_M/M)
} 
and that $LH_M/M$ is an abelian extension of degree $[L:~M]\ h(M)$.
Let $S$ denote the set of prime ideal $\p$ in $M$ such that 
 $\p$ is principal and $(\p, L/M)$ has order $f$.
Fix an element  $\s\in\Gal(L/M)$ having order $f$. 
Since $L/M$ is cyclic, we know that $\s^k,\ k\in (\ZZ/f\ZZ)^\t$ are exactly all the element in
$Gal(L/M)$ having order $f$. 
Moreover, we can interpret  the constraint that  $\p$ is principal as $(\p, H_M/M)=1$.
Under the isomorphism \eqref{eq_gal_split}, we know that 
\eqn{
S=\set{ \p \text{ in } M | (\p, LH_M/M) = (\s^k, 1)\in \Gal(L/M)\t\Gal(H_M/M),\ k\in (\ZZ/f\ZZ)^\t}.
}
A direct application of Proposition \ref{prop_density} yields
\eqn{
\delta(S) = \frac{\varphi(f)}{[L:M]\ h(M)}.
}

Let $S_1$ be  the set of primes of $M$ having relative degree one over $\QQ$.
An elementary argument (c.f. \cite[Section $4.6$, $(4.6.2)$]{janusz} tells us that 
$\delta(S\cap S_1) = \delta(S)$. 
Let $\p\in S\cap S_1$ and $p=\p\cap\ZZ$.
Since $M/\QQ$ is Galois, $p$ splits completely in $M$ and
every $\p'$ in $M$ lying over $p$ is also principal.
Moreover, the assumption that $L/\QQ$ is Galois ensures that all $(\p', L/M)$ are conjugate 
and hence  having  the same order $f$.
It follows that \eqn{
\delta(S(f, L/M)) = \frac{1}{[M:\QQ]}\ \delta(S\cap S_1) =  \frac{\varphi(f)}{[L:\QQ]\ h(M)}.
}
The proof is complete.
}

The following lemma gives a lower bound for the density of  $Q_p$.
\lemm{\label{lem_density}
Let    $p\equiv5\pmod 8$ be a prime and $Q_p$ defined by \eqref{eq_qp}.
Then we have \eqn{
\delta(Q_p) \ge  
 \frac{\varphi((p-1)/4)}{p-1} \left( \frac{1}{h(F_p)}-\frac{1}{h(E_p)} \right),
}}
\pf{
Clearly $K$ and $E_p$ are both Galois over $\QQ$.
Let $q\in Q_p$ and $\Q$ be any prime in $E_p$ lying over $q$.
Since $E_p$ is the decomposition field of $q$ in $K$, 
$q$ splits completely in $E_p$. Thus we have 
$(\Q, K/F_p)=(q, K/\QQ)$, which has order $(p-1)/4$. 
Applying Lemma \ref{lem_density_prin} we obtain \eqn{
\delta(S(\frac{p-1}{4}, K/E_p)) = \frac{\varphi((p-1)/4)}{(p-1)h(E_p)}.
}
A similar analyze for $K/F_p$ yields \eqn{
\delta(S(\frac{p-1}{4}, K/F_p)) = \frac{\varphi((p-1)/4)}{(p-1)h(F_p)}.
}
In view of $Q_p =  S(\frac{p-1}{4}, K/F_p) \setminus S(\frac{p-1}{4}, K/E_p)$,
we have \aln{
\delta(Q_p) &= \delta(  S(\frac{p-1}{4}, K/F_p) \setminus S(\frac{p-1}{4}, K/E_p) )\\
	&\ge	\delta(  S(\frac{p-1}{4}, K/F_p) ) -   \delta(  S(\frac{p-1}{4}, K/E_p) )\\
	&=		\frac{\varphi((p-1)/4)}{p-1} \left( \frac{1}{h(F_p)}-\frac{1}{h(E_p)} \right),
}
where the second line is due to the fact that 
we can sum the densities of two disjoint sets, which is easily seen by the definition.
So we finish the proof.
}

Our next goal is to show that if $p>p_0$ for some $p_0$,  the density $\delta(Q_p)$ is positive.
Recall that  $E_p\subseteq K=\QQ(\zeta_p)$ contains $F_p=\QQ(\sqrt{p})$. 
Since $\ord_p(q)=(p-1)/4$ is odd, the complex conjugation 
does not fix $E_p$.
It follows that $E_p$ is a totally imaginary cyclic extension of $\QQ$, and hence a CM-field
with $E_p^+=F_p$ being the maximal real subfield.
We write  $h_p$ for $h(E_p)$, and $h_p^+$ for $h(E_p^+)=h(F_p)$.
Thus from Proposition \ref{prop_cm} (a) we know that $h_p=h_p^+h_p^-$ for a positive integer $h_p^-$, 
which is the relative class number for $E_p$.
We now consider the asymptotic behaviour of $h_p^-$.  
\lemm{\label{lem_h-ge1}
With the previous notation we have \eqn{
\log h_p^- \ge \frac{1}{2}(\log p)(1+o(1)) \quad \text{ as $p\rightarrow \infty$}.
}}
\pf{
We follow the  same  method in \cite[Section 4]{Washington}. But the case here is simpler.
Let $U_p, W_p, R_p$ and $d_p$ be the  unit group, group of roots of unity,
 regulator and discriminant of $E_p$ respectively,
and let  $U_p^+, R_p^+$ and $d_p^+$ denote the corresponding objects for $E_p^+$. 
The ideal is to use Brauer-Siegel theorem (Proposition \ref{prop_cm} (d)) for $E_p/E_p^+$.
To verify the assumption of the theorem, we need to estimate the discriminants of $E_p^+$ and $E_p$.
Recall that $E_p^+=F_p=\QQ(\sqrt{p})$ and 
clearly we know that $d_p^+=p$. Then the relative discriminant formula (c.f. \cite[pp. 82]{gtm110})
gives \eqn{
|d_p|=N_{E_p/\QQ}(\mathcal D(E_p/E_p^+)) |d_p^+|^{[E_p:E_p^+]},
}
where $\mathcal D(E_p/E_p^+)$ is the relative different, which is a integral ideal in $\o_{E_p}$.
Thus we have \eq{\label{eq_dp}
|d_p|\ge  |d_p^+|^{[E_p:E_p^+]} =p^2.
}
Since the we have $[E_p:\QQ] = 2[E_p^+:\QQ]=4$ for all $p$, 
we know that \eqn{
\frac{[E_p:\QQ]}{\log|d_p|}\rightarrow 0\quad\text{ and }\quad
\frac{[E_p^+:\QQ]}{\log|d_p^+|}\rightarrow 0
}
and Brauer-Siegel theorem applies.
It follows that \aln{
\log(h_p R_p)&= \frac{1}{2}\log d_p + o(\log d_p)\\
\text{and } \log(h_p^+ R_p^+)&= \frac{1}{2}\log d_p^+ + o(\log d_p^+)
}
By Proposition \ref{prop_cm} (b) and (c) we have \eqn{
\log \left( \frac{R_p}{R_p^+} \right) = O(1).
}
Hence, noting that $\log d_p^+\le \frac{1}{2} \log d_p$ by \eqref{eq_dp}, 
we have \aln{
\log h_p^- &=  \log(h_p R_p) - \log(h_p^+ R_p^+) - \log \left( \frac{R_p}{R_p^+} \right)\\
&= \frac{1}{2}\log d_p - \frac{1}{2}\log d_p^+ + o(\log d_p)  + O(1)\\
&\ge \frac{1}{2}\log d_p - \frac{1}{4}\log d_p + o(\log d_p)\\
&= \frac{1}{4}(\log d_p)(1+o(1))\\
&\ge \frac{1}{2}(\log p)(1+o(1)).
}}
\prop{\label{prop_infty}
Let notation be as before,  $p\equiv5\pmod 8$ a prime and $Q_p$ defined by \eqref{eq_qp}.
Then we have 
\itmz{
\itm{a} the equation
\eq{\label{prop_infty.eq_beta=q}
\beta\bar\beta=q,\quad \beta\in\o_{E_p}
}
 has no solution for all  $q\in Q_p$;
\itm{b} if $h_p>h_p^+$ then the set $Q_p$ is infinite;
\itm{c} there exists a lower bound $p_0$ such that
if $p>p_0$, then the set $Q_p$ is infinite.
}
}
\pf{
Let $p\equiv5\pmod 8$ and $q\in Q_p$.
Recall that $\ord_p(q)=(p-1)/4$ is  odd 
and $E_p$ is the decomposition field of $q$ in $K=\QQ(\zeta_p)$ 
with $[E_p:\QQ]=4$,
so we have the prime decomposition
\eqn{
q\o_{E_p}=\Q_1\Q_2\Q_3\Q_4.
}
It is also noted before that
 the complex conjugation is not in the decomposition group of $q$.
Thus we may assume $\Q_3=\bar\Q_1$ and $\Q_4=\bar\Q_2$.

Now we assume that the equation \eqref{prop_infty.eq_beta=q} has solution
$\beta\in\o_{E_p}$, so  we have 
\eqn{
\beta\bar\beta\o_{E_p}=q\o_{E_p}=\Q_1\Q_2\bar\Q_1\bar\Q_2.
}
It follows that  the only possible decompositions of $\beta$ are 
\eq{\label{prop_infty.eq_beta_dec}
\beta\o_{E_p} = \Q_1\Q_2,\ \Q_1\bar\Q_2,\ \bar\Q_1\Q_2\text{ or }\bar\Q_1\bar\Q_2.
}
Write  $\Gal(E_p/\QQ)=\left<\s\right>$ with $\s$ of order $4$.
It follows that we can assume
\eqn{
\Q_1^{\s^t}=\Q_{t+1},\quad t=0,1,\dots,3.
}
Then \eqref{prop_infty.eq_beta_dec} tells us that \eqn{
1=\Q_1^{1+\s},\ \Q_1^{1+\s^3},\ \Q_1^{\s^2+\s}\text{ or }\Q_1^{\s^2+\s^3}\text{ in $Cl(E_p)$.}
}
Correspondingly, rising to the power to $1-\s,\ 1-\s^3,\ \s^2-\s\text{ or }\s^2-\s^3$,
we obtain the same equation \eq{\label{eq_q^(10s2)=1}
\Q_1^{1-\s^2}=1\text{ in }Cl(E_p).
}

On the other hand, by the definition \eqref{eq_qp}, we know that
there is a prime ideal $\Q$ in $E_p$  over $q$ such that 
$\Q$ is not principal while $\q=\P\cap \o_{F_p}$ is principal.
With loss of generality, we may assume that $\Q_1=\Q$.
Since $\q$ is principal, so is  $\q\o_{E_p}=\Q_1\bar\Q_1$,
which means that \eqn{
\Q_1^{1+\s^2}=1\text{ in }Cl(E_p).
}
Combining with \eqref{eq_q^(10s2)=1}, we have $\Q_1^2=1$ in $Cl(E_p)$.
However, since $p$ is the unique  finite rational prime that ramifies in $E_p/\QQ$,
Lemma \ref{lem_cm_parity} tells us that $h(E_p)$ is odd.
It follows that $\Q_1=1$ in $Cl(E_p)$, which is an contradiction because 
 $Q_1$ is not principal by the previous argument. 
Thus the  assumption we made before is false and we complete the proof for (a). 

Next, let  $p\equiv5\pmod 8$ and suppose that $h_p>h_p^+$.
 By Lemma \ref{lem_density}, we know that 
\eqn{
\delta(Q_p) 
\ge \frac{\varphi((p-1)/4)}{p-1} \left( \frac{1}{h_p^+}-\frac{1}{h_p} \right) > 0.
}
Thus $Q_p$ is a infinite set and (b) is correct.

For the last assertion (c), recall that $E_p/F_p$ is CM  and 
 we use Lemma \ref{lem_h-ge1} to obtain \eqn{
h_p^- \rightarrow \infty  \quad \text{as $p\rightarrow \infty$}.
}
It follows that  there exists a lower bound $p_0$ such that
if $p>p_0$, then $h_p^->1$, i.e. $h_p>h_p^+$.
It follow by (b) that $Q_p$ is infinite for all $p>p_0$. 
That's all the proof.
}
Although we have shown that $Q_p$ is infinite, one do not know whether 
a given $q$ is in $Q_p$. 
However,  this can be done when we know the Hilbert class field $H_{F_p}$ and 
$H_{E_p}$ of $F_p$ and $E_p$, respectively. We now describe this method as follows.

Suppose in  general, $L$ is a number field. Let $\Xi_L(x)\in\o_L[x]$
be an irreducible polynomial having a root that generates the Hilbert 
class field $H_L$  over $L$, and we call $\Xi_L$ the  \emph{Hilbert class polynomial}.
If there is a subfield $M$ such that   $L/M$ is  cyclic  with $h(M)=1$, then there exist an auxiliary
field  $K=K_{L/M}$ such that
$H_L=KM$ and $M=K\cap M$. See \cite[Proposition 3]{splt.hilb}.
It follows that  we can choose $\Xi_L(x)$ with coefficients in  $\o_M$.
\lemm{\label{lem_prin_by_poly}
Let $L/M$,  $\Xi_L(x)\in \o_M[x]$ be as before, 
and $\P$  a prime ideal of $L$  having relative degree one over $M$ not dividing the discriminant of $\Xi_L$.
Let $\p=\P\cap\o_M$.
Then $\P$  is principal if and only if 
$\Xi_L(x)=0$ has a solution over $\o_M/\p$.
}
\pf{
Since $H_L$ is the Hilbert class field of $L$, so we know that $\P$ is principal if and only if
$\P$ splits completely in $H_L$.
Note that $\P$ does not  divide the discriminant of $\Xi_L$.
Then  a direct application of  Kummer theorem (c.f. \cite[Proposition 25, Chapter I.8]{gtm110})
tells us that $\P$ splits completely in $H_L$ if and only if 
$\Xi_L(x)=0$ has a solution over $\o_L/\P$, 
which is to say that 
$\Xi_L(x)=0$ has a solution over $\o_E/\p$, 
since 
$\P$ is a prime ideal of $L$  having relative degree one over $M$.
}
In our case where $E_p/\QQ$ is cyclic and $F_p$ is real quadratic, we know that 
both $K_{E_p/\QQ}$ and $K_{F_p/\QQ}$ exist. Hence we have $\Xi_{E_p}$ and $\Xi_{F_p}$ with integral coefficients.
Then we have
\cor{\label{cor_qp_by_poly}
Fix a prime $p\equiv 5\pmod8$ and let the irreducible polynomials $\Xi_{E_p}(x)$ and $\Xi_{F_p}(x)$ 
in $\ZZ[x]$  be as before.
Given $q\in \tilde Q_p$ not dividing the discriminants of the two polynomials, then we have
  $q\in Q_p$ if and only if   $\Xi_{F_p}(x)\equiv0 \pmod q$ is  solvable while
  $\Xi_{E_p}(x)\equiv0 \pmod q$ is not.
}
\pf{
Apply Lemma \ref{lem_prin_by_poly} to $E_p/\QQ$ and $F_p/\QQ$ 
and then the result follows from the definition of $Q_p$.
}
Now the problem left is to find the Hilbert class polynomials for $E_p$ and $F_p$.
For the real quadratic field $F_p$, the polynomial $\Xi_{F_p}$ is quite easy to obtain
(c.f. Stark’s method described by Cohen \cite{gtm193}, who also gives a list of these polynomials).
As for $E_p$, which is   a CM-field and is cyclic of degree $4$ over $\QQ$,
we could use complex multiplication to calculate $\Xi_{E_p}$.
This is more complicated than the imaginary quadratic case where elliptic curves and the $j$-invariant 
are enough.
In the case of degree $4$ CM-fields, we work with curves of genus $2$, three $j$-invariants
and three igusa class polynomials. See Streng \cite{streng2014computing} or
Enge et al. \cite{enge2013computing} for  complete description of the method.
There is also an implementation of the algorithm by Enge et al., CMH \cite{cmh-igusa},
which enables us to calculate the individual igusa class polynomials.
We can use igusa class polynomials instead of the Hilbert class polynomial,
or calculate the Hilbert class polynomial by them.
This solves the whole problem.
We give an example  to illustrate this method.
\eg{ Let $p=101$, and 
\aln{
\Xi(x)\ =\ x^5\ -\ 1237224274356339549352800\ x^4\ +\ 57176933499148\qquad&\\
833882237435031573248869838360576\ &x^3\\
 +\ 2514056979190981026432576749022147825857609219676093\qquad&\\
86630877092098080768\ &x^2\\
 -\ 1023671146645480759972364788108250129958692705554245\qquad&\\
7706457967352624977868378319649505280\ &x\\
 +\ 1530499568113365603805244886351320629567046080073893\qquad&\\
31920814045884605474516662499805087162052695075848192.&
}
If $q\in \tilde Q_p$ and $q \neq 2542000616863$, then $q\in Q_p$ if and only if 
$\Xi(x)\equiv0 \pmod q$ is solvable.
}
\pfs{
For $p=101$ we have $F_p=\QQ(\sqrt{101})$ and $E_p=\QQ(\alpha)$
 where $\alpha$ is a root of $x^4+101x^2+101$.
Using GP calculator (see \cite{pari}) we obtain that 
$h_p^+=h(F_p)=1$ and $h(E_p)=5$. 
Using CMH (see above), we obtain that the first  igusa class polynomial of $E_p$ is
$\Xi(x)$, which is also the Hilbert class polynomial of $E_p$ since it already has degree $5$.
Also we know $q=2542000616863$ is the only prime in $\tilde Q_p$ that divides the discriminant of $\Xi(x)$.
The assertion  follows by Corollary \ref{cor_qp_by_poly}.
}

Let us turn to the 
\pf{[Proof of Theorem \ref{thm_p5mod8}]
If there exist PPS  with type $[p, n]$, 
where $p\equiv5\pmod 8$ be a prime and  $ n=p^aqn'$,
then by Proposition \ref{prop_eq_pps} we know that
\eqn{
\alpha\bar\alpha=n=p^aqn'\text{ for some }\alpha\in\o_K=\ZZ[\zeta_p].
}
Here $K=\QQ(\zeta_p)$.
\cite[Lemma 2.4 (2)]{liu2016new} tells us that 
\eqn{
\alpha_1\bar\alpha_1=p^aq\text{ for some }\alpha_1\in\o_K.
}
Next by \cite[Lemma 2.4 (1)]{liu2016new} we obtain that
\eqn{
\alpha_2\bar\alpha_2=q\text{ for some }\alpha_2\in\o_K.
}
We may assume $p>5$. Then $\ord_p(q)=(p-1)/4>1$ and so $(p,q-1)=1$.
Recall that $E_p$ is the decomposition field of $q$ in $K$,
so we use \cite[Lemma 2.4 (3)]{liu2016new} to obtain that
\eqn{
\beta\bar\beta=q\text{ for some }\beta\in\o_K\text{ and }\beta^2\in\o_{E_p}.
}
But $[K:E_p]=\ord_p(q)=(p-1)/4$ is  odd, 
so in fact we have $\beta\in\o_{E_p}$.

Thus the theorem follows from Proposition \ref{prop_infty}.
}

\section{Non-existence result for PPSs with type $[p\equiv3\pmod 4,p^aq^ln']$ for certain $p, q$ and $l$}\label{sec_sp}
We will prove Theorem \ref{thm_sp} in this section.
The main ideal is the application of Stickelberger relations,
which was used by the first author and Jianing Li \cite{lv2016new} for
showing  non-existence results for some bent functions.
First we fix some additional notation. 
Suppose $n=p^aq^ln'$ and $p\equiv3 \pmod 4$ be as in the theorem.
Then we know that  $f:=\ord_p(q)$ is odd. Thus $g:=\frac{\varphi(p)}{f}$ is even and we set $u=g/2$.
Recall that $K = \QQ(\zeta_p)$ and 
let $E$ be the unique subfield of $K$ having degree $g$ over $\QQ$.
Then  $E$ is the decomposition group of $q$ in $K$ and
 is CM with $E^+=E\cap \RR$ its maximal real subfield 
 (the argument is similar as before, see  Section  \ref{sec_p5mod8}).
It is well known that  $K$ contains the unique 
imaginary quadratic subfield $F=\QQ(\sqrt{-p})\subset E$.

Suppose the  prime decomposition of $q$ in $E$ is
 \eq{\label{eq_q_dec}
q\o_E=\Q_1\Q_2\dots\Q_g.
}  
If there is a  PPS  with type $[p,n]$, then 
as in the proof of Theorem \ref{thm_p5mod8},
a similar argument using  Proposition \ref{prop_eq_pps} and 
\cite[Lemma 2.4]{liu2016new} yields the equation
\eqn{
\beta\bar\beta=q^l,\ \beta\in\o_E.
}
Since  $f$ is odd, the complex conjugation is not in the decomposition group of $q$.
Thus we may assume $\Q_{u+k}=\bar\Q_k,\ k=1,2,\dots,u$.
Then we have \eqn{
\beta\bar\beta\o_E=\prod_{k=1}^u\Q_k^l\bar\Q_k^l.
}
So \eq{\label{eq_beta_dec}
\beta\o_E=\prod_{k=1}^u\Q_k^{l_k}\bar\Q_k^{\bar l_k}
}
 where $l_k, \bar l_k$ are nonnegative integer such that 
$l_k+\bar l_k=l$ for all  $k=1,2,\dots,u$.

For convenience we write $x_k$ for $\Q_k$ in $Cl(E)$ and view $Cl(E)$ additively. 
Hence \eqref{eq_beta_dec} becomes
\ga{
\sum_{k=1}^u(l_k x_k + \bar l_k \bar x_k) = 0  \label{eq_orig_rel} \\
\text{where }l_k+\bar l_k=l,\ k=1,2,\dots,u.\nonumber
}
Thus we obtain the 
\prop{\label{prop_sp_general}
With the above notation, if  \eqref{eq_orig_rel}
has  no  nonnegative integral solution 
$(l_1,l_2,\dots,l_g)$, where
$l_k+\bar l_k=l_0$ and  $l_{u+k}:=\bar l_k,\ k=1,2,\dots,u$,
then there is no PPS with type $[p,n]$.
}

To show  \eqref{eq_orig_rel} is not solvable in the above sense, 
we  have to exploit the relations between $x_k$'s in $Cl(E)$.
By \eqref{eq_q_dec} we have
\eq{\label{eq_sum_x}
\sum_{k=1}^g x_k=0.
}
We want to find more relations.

Let  $K^+=K\cap\RR=\QQ(\zeta_p+\zeta_p^{-1})$.
Then  Miller's work on class number of $K^+$ gives
\thmu{[\cite{miller2015real}, Theorem $1.1$]
The class number of 
$\QQ(\zeta_p+\zeta_p^{-1})$ is $1$  if $p\le151$ is a prime.
}

From now on we  suppose   $p\le 151$.
Clearly, $E^+\subseteq K^+$. Then 
by Miller's result and Lemma~\ref{lem_h_div}, we have
$h(E^+)=h(K^+)=1$.
Now   $q\o_{E^+}=\q_1\q_2\dots\q_u$ where $\q_k\o_E=\Q_k\bar\Q_k$,
 and all $\q_k$'s are principal since  $h(E^+)=1$.
This implies  the relations
\eq{\label{eq_conj_x}
x_k+x_{u+k}=0,\quad k=1,2,\dots,u.
}

However, these relations above are not enough. We need  the
Stickelberger ideal introduced in Section~\ref{sec_facts}.
Let $\Q=\Q_1$ and correspondingly $x=x_1$.
Let $c$ be an integer not divisible by $p$. 
Since it is well-known that $p$ is the minimal integer such that $F\subseteq\QQ(\zeta_p)$,
it follows that $p$ is also the minimal one such that $E\subseteq\QQ(\zeta_p)$.
By  Proposition~\ref{prop_stick}, we have 
\eq{\label{eq_orig_ann}
(c-\s_c)\theta\ \Q=1\text{ in }Cl(E).
}
Let $w$ be a primitive root mod $p$. Recall  $G=\Gal(K/\QQ)$. Then  the decomposition group of $q$ in $K$ is
$\left<q\right>=\left<w^g\right>\subseteq G=(\ZZ/p\ZZ)^\t$. It follows that we can assume
\eq{\label{eq_sigma_x}
\s_w^{tg+s}(x)=x_{s+1},\quad t\in\ZZ,\ s=0,1,\dots,g-1.
}
Let $k_{c,a}=[\frac{ca}{p}]$ for any integer $a$. We have \aln{
(c-\s_c)\theta&=(c-\s_c)\sum_{a\in(\ZZ/p\ZZ)^\t}\left\{\frac{a}{p}\right\} \s_a^{-1} \\
&=\sum_a \left(c\left\{\frac{a}{p}\right\}-\left\{\frac{ca}{p}\right\}\right)\s_a^{-1}\\
&=\sum_{a=1}^{p-1}k_{c,a}\s_a^{-1} \quad\text{ (the definition of $k_{c,a}$) }\\
&=\sum_{s=0}^{p-2}k_{c,w^{-s}}\s_w^s \quad\text{ ($w^{-s}$ means $w^{-s} \mod p$) }\\
&=\sum_{t=0}^{f-1}\sum_{s=0}^{g-1}k_{c,w^{-(tg+s)}}\s_w^{tg+s}
}
Then by \eqref{eq_orig_ann} we have
\aln{
1&=\Q^{\sum_{t=0}^{f-1}\sum_{s=0}^{g-1}k_{c,w^{-(tg+s)}}\s_w^{tg+s}},\\
\text{ i.e., }\quad
 0&=\sum_{t=0}^{f-1}\sum_{s=0}^{g-1}k_{c,w^{-(tg+s)}}\s_w^{tg+s}(x)\\
&=\sum_{t=0}^{f-1}\sum_{s=0}^{g-1}k_{c,w^{-(tg+s)}}x_{s+1} \quad\text{ (by \eqref{eq_sigma_x})}\\
&=\sum_{s=1}^g m_{c,s} \sum_{t=0}^{f-1}k_{c,w^{-tg-s+1}} x_s.
}
If we set \eq{\label{eq_m_cs}
m_{c,s}= \sum_{t=0}^{f-1}k_{c,w^{-tg-s+1}}
}
we have $p-1$ linear equations
\eqn{
\sum_{s=1}^g m_{c,s}x_s = 0,\quad c=1,2,\dots p-1.
}
We now combine these $p-1$ equations, together with the equation \eqref{eq_sum_x} and
the $u$ equations \eqref{eq_conj_x}, to give a whole   collection of equations 
\eq{\label{eq_ann}
XM_{p,f}^T=0
}
 where $M_{p,f}$ is a $(p+u)\t g$ matrix with integer entries made of 
the coefficients of all  the $p+u$ equations and $X=(x_1, x_2,\dots, x_g)$.
Note that $M_{p,f}$ depends only on $p$ and $f$.
To simplify these relations of $x_1,x_2, \dots,x_g$, we need to calculate  the 
\emph{Hermite normal form} of $M_{p,f}$. By the well-known result 
(c.f. \cite[Chapter~2.4.2]{Cohen}) for the existence of the Hermite normal form,
there exists a unique matrix  $U_{p,f}\in\GL_{p+u}(\ZZ)$, such that $H_{p,f}=M_{p,f}^TU_{p,f}$ is a Hermite normal form.
It follows from \eqref{eq_ann} that 
\eqn{
XH_{p,f}=0.
}
In fact, $H_{p,f}$ can be obtained by applying  a finite sequence of elementary row operations over 
$\ZZ$ from $M_{p,f}^T$.

Now with the help of a computer and
using a simple program or a computer algebra system,
 we can calculate the individual Hermite normal form $H_{p,f}$ for
\eqn{
(p, f)\in \set{(31 , 5),\ (127, 9),\ (127, 21),\ (139, 23),\ (151, 15)}.
}

Let us take $(p,f)=(31,5)$ and $(151,15)$ for example.
Thus we obtain the relation
\eq{\label{eq_final_rel_31}
(x_1, x_2,x_3)\begin{pmatrix}
18&8&15\\
0&2&1\\
0&0&1\end{pmatrix}=0
}
for $(p,f)=(31,5)$ and 
\eq{\label{eq_final_rel_151}
X_{151}\begin{pmatrix}
3934&1304&3470&3544&1477\\
0&2&0&0&1\\
0&0&2&0&1\\
0&0&0&2&1\\
0&0&0&0&1\end{pmatrix}=0
}
for $(p,f)=(151,15)$,
where $X_{151}:=(x_1, x_2,\dots,x_5)$ and 
we omit $x_{u+1},\dots,x_g$ and other parts of $H_{p,f}$
since  $x_{u+k}=-x_k$.

Using these computational results, we can turn to the
\pf{[Proof of  Theorem \ref{thm_sp}]
We have to verify the assumption in Proposition \ref{prop_sp_general}.
\enmt{
\it If $(p,f)=(31,5)$ the first column of the matrix in \eqref{eq_final_rel_31} tells us 
that $18x_1=0$ in $Cl(E)$. 

Recall that  $K=\QQ(\zeta_p), h_p=h(K)$
and $h_p^+=h(\QQ(\zeta_p+\zeta_p^{-1}))=1$. 
We can write  $h_p = h_p^+ h_p^-$ (see Proposition \ref{prop_cm} (a)).
By \cite[Tables~\S3, pp. 412-420]{Washington} we know $h_{31}^-$ is odd.
Since  $h(E)\mid h(K)=h_p=h_p^-$, we know $h(E)$ is also odd.
It follows that $9x_1=0$ and $\ord(x_1)=1,3$ or $9$ in $Cl(E)$.

We claim that $\ord(x_1)=9$.
Recall   $F=\QQ(\sqrt{-p})=\QQ(\sqrt{-31})\subseteq E$  and let $\q_F=\Q_1\cap\o_F$.
By the table in \cite[Section 12.1.2]{gtm193} we know that $\Xi_{31}(x)$ is the  Hilbert
class polynomial of $F$.
Thus  $h(F)=\deg(\Xi_{31}(x))=3$ and
the same argument as in the proof of 
Lemma \ref{lem_prin_by_poly}
tells us that
$\q_F$ is not principal if and only if 
$\Xi_{31}(x)\equiv0\pmod q$ is not solvable.
By the assumption in  Theorem \ref{thm_sp} we know this is the case and
then 
 $\q_F$ has order $3$ in $Cl(F)$.
If $\ord(x_1)=1$, i.e. $\Q_1=1$ in $Cl(E)$, then taking norm gives $\q_F=1$ in $Cl(F)$,
which is a contradiction.
If $\ord(x_1)=3$, then $\left<x_1\right>\cong\ZZ/3\ZZ$  and we may assume 
$x_1=1\mod3$. The second column of the matrix reads $8x_1+2x_2=0$.
Since $2$ can be canceled from every equation, we have 
 $x_2=-4x_1$. Hence $x_2=-x_1=-1\mod3$ and similarly $x_3=1\mod3$.
Thus $x_k=\pm1\mod3\in\ZZ/3\ZZ$ for all $k=1,2,\dots,6$.
But we know three of all six $x_k$'s (i.e. $\Q_k$'s) lie over $\q_F$.
Suppose that $\q_F\o_E = \Q_{k_1} \Q_{k_2} \Q_{k_3}$.
If all these three $x_{k_1},x_{k_2},x_{k_3}$ are the same, say $1\mod3$, then 
$\q\o_E=1$ in $Cl(E)$.
Since  $Cl(F)\lr Cl(E)$  is injective (Proposition~\ref{prop_cl_inj}),
we have $\q_F=1$ in $Cl(F)$, a contradiction.
Otherwise we may assume $x_{k_1}=-x_{k_2}=1$ and then $x_{k_1}+x_{k_2} = 0$.
 Taking norm gives $\q_F^2=1$, which
is also false.

Thus we have $\ord(x_1)=9$ and using the matrix again we obtain
\eqn{
(x_1,x_2,\dots,x_6)=(1,-4,-2,-1,4,2)
}
 are all in $\left<x_1\right>\cong\ZZ/9\ZZ$.
We now apply Proposition \ref{prop_sp_general}.
Let $l=1,3\dots$ and solve the equation  \eqref{eq_orig_rel} modulo $9$.
A simple calculation tells us that $l_0=1$
is the maximal nonnegative odd number such that 
\eqref{eq_orig_rel} is not solvable in $Cl(E)$.
Hence we obtain by 
Proposition~\ref{prop_sp_general}
 the non-existence of GBFs with type $[31, 31^aq^ln']$.

\it The argument  for $(p,f)=(151,15)$ is similar. Using the matrix in \eqref{eq_final_rel_151} 
we know that $2\t7\t281x_1=0$. 
The same method yields the fact  that $h(E)$ is also odd.
Thus we find that $\ord(x_1)=7,281$ or $1967$. 
In this case $F=\QQ(\sqrt{-157})$.
Knowing  that  $h(F)=7$ and  $\q_F$ has order $7$ in $Cl(F)$
since $\Xi_{151}(x)\equiv0\pmod q$ is not  solvable,
the candidate order $1$ and $7$ can be removed by the previous 
method. If we have $281x_1=0$, taking norm gives  $\q_F^{281}=1$,
which contradicts to $\ord(\q_F)=7$. Thus   $\ord(x_1)=1967$ and  we obtain
$x_1,\dots,x_{10}\in\left<x_1\right>\cong\ZZ/1967\ZZ$ and
\gan{
(x_1,x_2,\dots,x_5)=( 1, -652, 232, 195, 715 )\\
x_{5+k}=-x_k,\ k=1,2,\dots,5.
}
Let $l=1,3\dots$ and solve the equation  \eqref{eq_orig_rel} modulo $1967$.
We find that $l_0=5$ 
is the maximal nonnegative odd number such that 
\eqref{eq_orig_rel} is not solvable in $Cl(E)$.
Again  Proposition~\ref{prop_sp_general}
implies  the non-existence of GBFs with type $[151, 151^aq^ln']$.

\it For other $(p,f)\in\set{(127, 9),\ (127, 21),\ (139, 23)}$,
the proofs are similar.
}
}

\section{Corresponding non-existence results for PAPSs}
In this section, we give  briefly
tow  non-existence results for PAPSs, which are similar to 
Theorem \ref{thm_p5mod8} and \ref{thm_sp}, respectively. Their proofs are also similar.
\propu{[See \cite{liu2016new} Theorem 1.4 (2)]\label{prop_eq_paps}
If there exist PAPS  with type $[p, n+1]$, then $p\mid n-1$ and 
$\alpha\bar\alpha=n$ for some $\alpha\in\ZZ[\zeta_p]$.
}

\thm{ 
Let $p\equiv5\pmod 8$ be a prime and $\tilde Q_p=\set{q\text{ is a prime } | \ord_p(q)=(p-1)/4}$.
Then there exists a lower bound $p_0$, and an infinite set $Q_p\subseteq \tilde Q_p$
for each $p$, such that
if $p>p_0$, there is no PAPSs with type $[p, qn'+1]$ for all integers $q\in Q_p$, $n'$ 
such that $n'=1$ or $\fracn{p'}{p}=-1$ for all prime divisor $p'$ of $n'$ and $p\mid qn'-1$.
}
\pfs{
If there exist PAPS  with type $[p, qn'+1]$, 
where $p\equiv5\pmod 8$ be a prime,
then by Proposition \ref{prop_eq_paps} we know that
$p\mid qn'-1$ and 
\eqn{
\alpha\bar\alpha=qn'\text{ for some }\alpha\in\o_K=\ZZ[\zeta_p].
}
Here $K=\QQ(\zeta_p)$.
By \cite[Lemma 2.4 (1)]{liu2016new} we obtain that
\eqn{
\alpha_2\bar\alpha_2=q\text{ for some }\alpha_2\in\o_K.
}
Then the remaining  argument  is totally the same as the proof of 
Theorem \ref{thm_p5mod8}. See Section \ref{sec_p5mod8}.
}

\thm{
Let $p\equiv3$ be a prime, $q\neq p$ another prime  and $f=\ord_p(q)$.
Suppose that the triple $(p, f, l_0)$ equals to one of the following value:
\eqn{
(31 , 5 , 1),\ (127, 9 , 1),\ (127, 21, 3),\ (139, 23, 1),\ (151, 15, 3).
}
Define \aln{
\Xi_{31}(x)					&=  x^3 + x - 1,\\
\Xi_{127}(x)					&=  x^5 - x^4 - 2x^3 + x^2 + 3x - 1,\\
\Xi_{139}(x)					&=  x^3 - x^2 + x + 2,\\
\text{and}\quad \Xi_{151}(x)	&=  x^7 - x^6 + x^5 + 3x^3 - x^2 + 3x + 1.
}
Suppose further that for each $p\in\set{31,127,139,151}$, the corresponding $q$ satisfies that
$\Xi_p(x)\equiv0\pmod q$ is not  solvable.
Then there is no PPPs with type $[p, q^ln'+1]$ for all integers $l$ odd, $1\le l\le l_0$,
$n'$ such that $n'=1$ or $\fracn{p'}{p}=-1$ for all prime divisor $p'$ of $n'$ and
$p\mid q^ln'-1$.
}
\pfs{
The  argument  is totally the same as the proof of 
Theorem \ref{thm_sp} (in Section \ref{sec_sp}), except that we
use  Proposition \ref{prop_eq_paps} instead of Proposition \ref{prop_eq_pps}.
}

\section*{Acknowledgment}
The author would like to thank  Jianing Li  for many helpful discussions and comments.

%bib
\bibliography{\bibfilename}
\bibliographystyle{amsplain} 
\end{document}